\documentclass[conference]{IEEEtran}

\usepackage[utf8]{inputenc}
\usepackage{cite}
\usepackage{amsmath}
\usepackage{amssymb}
\usepackage{mathtools}
\usepackage{dblfloatfix}
\usepackage{multirow}
\usepackage{mathdots}
\usepackage{bm}
\usepackage{slashbox}

\newcommand\w[1]{\makebox[1.8em]{$#1$}}

\ifCLASSINFOpdf
  \usepackage[pdftex]{graphicx}
  \graphicspath{{./pdf/}}
  \DeclareGraphicsExtensions{.pdf}
\else
  \usepackage[dvips]{graphicx}
  \graphicspath{{./eps/}}
  \DeclareGraphicsExtensions{.eps}
\fi

\begin{document}

\title{Frequency-Domain Modeling of OFDM Transmission with Insufficient Cyclic Prefix using Toeplitz Matrices}


\author{\IEEEauthorblockN{Grzegorz Cisek and Tomasz P. Zieliński}
\IEEEauthorblockA{AGH University of Science and Technology, Department of Telecommunications, Cracow, Poland}
E-mail: \textit{gcisek@agh.edu.pl}}

\maketitle

\begin{abstract}

A novel mathematical framework is proposed to model Intersymbol Interference (ISI) phenomenon in wireless communication systems based on Orthogonal Frequency Division Multiplexing (OFDM) with or without cyclic prefix. The framework is based on a new formula to calculate the Fast Fourier Transform (FFT) of a triangular Toeplitz matrix, which is derived and proven in this paper. It is shown that distortion inducted by the ISI from a given subcarrier is the most significant for the closest subcarriers and the contribution decays as the distance between subcarriers grows. According to numerical experiments, knowledge of ISI coefficients concentrated around the diagonal of Channel Frequency Response (CFR) matrix improves the receiver's error floor significantly. The potential use of the framework for real-time frequency domain channel simulation was also investigated and demonstrated to be more efficient than conventional time domain Tapped Delay Line (TDL) model when a number of simulated users is high.

\end{abstract}

\begin{IEEEkeywords}
OFDM, Multipath channels, Intersymbol Interference, Interblock Interference, Simulation, Toeplitz matrix.
\end{IEEEkeywords}

\section{Introduction}

Intersymbol Interference (ISI) in OFDM-based wireless communication systems occurs when duration of a cyclic prefix is not long enough to compensate an influence of multipath channel. The cyclic prefix is added at the beginning of each transmitted data block to minimize the effects of ISI from the previous OFDM symbol and to allow representation of a linear convolution of transmit signal with Channel Impulse Response (CIR) as a circular convolution during frequency domain equalization. Cyclic prefix is detached at the receiver, but its duration may be insufficient to fully mitigate the effect of ISI introduced by multipath channel with very long delay spread. Because cyclic prefix is redundant and does not carry any useful information, it limits the overall capacity of the system.

One of the methods to reduce the distortion produced by insufficient cyclic prefix is interference estimation using pilot symbols and equalization at the receiver. In \cite{Parsaee04} and \cite{Nisar07}, channel equalizers based on MMSE criterion were proposed. Equalization based on iterative cancellation was studied in \cite{Molisch07}, whereas \cite{Pham17} dealt with a multistep estimation for MIMO systems. Another approach to reduce the negative impact of ISI is precoding performed at the transmitter. A method based on channel independent precoding with interference alignment was proposed in \cite{Jin13}. In \cite{Kim17}, transmit precoding was applied to reduce the RMS delay spread of the channel. Moreover, characterization and mitigation of ISI with either precoding or equalization is also an important aspect of baseband Discrete Multitone (DMT) systems \cite{Freire14}.

In this paper, a new framework to characterize ISI in OFDM systems is proposed. The framework originates from a commonly accepted matrix-based model of channel convolution \cite{Parsaee04,Nisar07,Jin13}, but it makes one step forward to outline structure and behavior of the interference in frequency domain. In comparison to the framework proposed by Wu \cite{Wu06}, a smoother approach is presented that directly describes ISI in term of dependence between OFDM subcarriers, cyclic prefix length and CIR.

Beyond potential application to the interference cancellation, estimation and equalization, the authors see a possibility to utilize the proposed framework in the area of frequency domain channel modeling: using it together with the Fast Fourier Transform of triangular Toeplitz matrix derived in the appendix. The existing frequency domain wideband channel models do not have a possibility to incorporate the effects of ISI \cite{Cisek17, Cisek18}. However, it is highly desired to provide such functionality in order to extend capability of existing simulators to cover high delay spread scenarios.

\section{Time Domain Model}

Multipath time- and frequency-selective fading channel applied to digital baseband signal \(x(n)\) can be implemented as a non-recursive filter with time variant impulse response:
\begin{equation}
\label{eqn_tdl_y}
y(n) = \displaystyle\sum_{m=-\infty}^{\infty} h(n,m) x(n-m) + w(n)
\end{equation}
where \(w \sim \mathcal{CN}(0,\sigma^2)\) is a zero-mean circularly symmetric complex Gaussian noise with variance \(\sigma^2\) and \(h(n,m)\) is given by:
\begin{equation}
\label{eqn_tdl_h}
h(n,m) = \displaystyle\sum_{l=0}^{L-1} c_l(n) \delta(m-d_l)
\end{equation}

Scalars \(d_l\) are excess delays of dominant channel taps \(l=0,\dots,L-1\) and function \(\delta(.)\) is the Kronecker delta. Time-varying coefficients \(c_l(n)\) are Rayleigh distribution and modeled as complex Gaussian random processes, correlated in time and statistically independent across different paths. The temporal autocorrelation function is given by:
\begin{equation}
\label{eqn_rayleigh_corr}
\mathbb{E}\{c_l(n) \cdot c_l(n+p)^\ast\} = a_l \cdot J_0 (2 \pi p f_D T_s)
\end{equation}
where \(J_0\) is the zero-order Bessel function of the first kind, \(a_l\) is an average path gain, \(f_D\) is a maximum Doppler frequency and \(T_s\) is a sampling interval.

Because the data are transmitted in time blocks, not continuously, (\ref{eqn_tdl_y}) can be rewritten to equivalent block-based equation to express transmission of \(u\)-th OFDM symbol. Vectors \(\bm{x_u}\) and \(\bm{y_u}\) denote data of \(x(n)\) and \(y(n)\) in \(u\)-th symbol before/after cyclic prefix removal/insertion. Moreover, denote \(N\) a FFT size of considered OFDM system, \(v\) a length of a cyclic prefix, \({\tau=\max(d_l)}\) a delay spread span of the channel and \(v_{is} = \tau - v\) a length of ISI term exceeding the cyclic prefix. 

The CIR matrix \(\bm{H}\) of size \(N \times N\) with sufficient cyclic prefix, i.e. \(v \ge \tau\), is given by:
\begin{equation}
\label{eqn_tdl_mat_h}
\bm{H} = \left[
 \begin{smallmatrix}
  h(0,0)       & 0      & \cdots    & h(0,\tau)   & \cdots & h(0,1)         \\
  \vdots       & \ddots & 0         & \cdots      & \ddots & \vdots         \\[0.5em]
  h(\tau,\tau) & \cdots & h(\tau,0) & 0           & \cdots & 0              \\
  \vdots       & \ddots & \ddots    & \ddots      & \ddots & \vdots         \\[0.5em]	   
  \w0          & \w\cdots & \w0     & h(N-1,\tau) & \w\cdots & h(N-1,0)
 \end{smallmatrix} \right]
\end{equation}

ISI components of the channel are denoted by matrices \(\bm{A}\) and \(\bm{B}\) of size \(N \times N\) to represent the overlap of marginal elements in upper triangular part of \(\bm{H}\) \cite{Jin13}. A sub-matrix \(\bm{S}\) of size \(v_{is} \times v_{is}\) corresponds to a part of CIR exceeding cyclic prefix duration. Matrices \(\bm{A}\), \(\bm{B}\), and \(\bm{S}\) can be given as:
\begin{equation}
\label{eqn_tdl_mat_a}
\bm{A} = \left[
 \begin{smallmatrix}
  \bm{0}_{v_{is} \times (N - \tau)}       & \bm{S}                               & \bm{0}_{v_{is} \times v} \\[0.5em]
  \bm{0}_{(N - v_{is}) \times (N - \tau)} & \bm{0}_{(N - v_{is}) \times v_{is}}  & \bm{0}_{(N - v_{is}) \times v}
 \end{smallmatrix} \right]
\end{equation}

\begin{equation}
\label{eqn_tdl_mat_b}
\bm{B} = \left[
 \begin{smallmatrix}
  \bm{0}_{v_{is} \times (N - v_{is})}       & \bm{S}   \\[0.5em]
  \bm{0}_{(N - v_{is}) \times (N - v_{is})} & \bm{0}_{(N - v_{is}) \times v_{is}}
 \end{smallmatrix} \right]
\end{equation}

\begin{equation}
\label{eqn_tdl_mat_s}
\bm{S} = \left[
 \begin{smallmatrix}
  h(0, \tau)  & h(0, \tau - 1)  & \cdots & h(0, v+1) \\[0.5em]
  0           & h(1, \tau)    & \cdots & h(1, v+2) \\[0.1em]
  \vdots      & \vdots            & \ddots & \vdots \\[0.4em]
  0           & 0                   & \cdots        & h(v_{is}-1, v+v_{is}) \\
 \end{smallmatrix} \right]
\end{equation}
where \(\bm{0}_{m \times n}\) denotes a null matrix of size \(m \times n\).

Data vector for symbol \(u\) received after passing the multipath fading SISO channel under insufficient cyclic prefix conditions, i.e. \(v < \tau\), is given by:
\begin{equation}
\label{eqn_tdl_mat_y_isi}
\bm{y_u} = (\bm{H} - \bm{A}) \bm{x_u} + \bm{B} \bm{x_{u-1}} + \bm{w_u}
\end{equation}

When no ISI occurs, i.e. \(v \ge \tau\), \(\bm{A}\) and \(\bm{B}\) are null matrices, (\ref{eqn_tdl_mat_y_isi}) simplifies to a case covered in \cite{Cisek17}:
\begin{equation}
\label{eqn_tdl_mat_y}
\bm{y_u} = \bm{H} \bm{x_u} + \bm{w_u}
\end{equation}

We propose to further rearrange (\ref{eqn_tdl_mat_y_isi}) in order to separate Intercarrier and Intersymbol Interference terms. Define a circular-shift permutation matrix of size \(N \times N\):
\begin{equation}
\label{eqn_P}
\bm{P} = \left[
 \begin{smallmatrix}
  0      & 0      & \cdots & 0      & 1     \\[0.5em]
  1      & 0      & \cdots & 0      & 0     \\
  0      & 1      & \ddots & 0      & 0     \\
  \vdots & \ddots & \ddots & \ddots & \vdots\\[0.3em]
  0      & \cdots & 0      & 1      & 0     \\
 \end{smallmatrix} \right]
\end{equation}

The matrices \(\bm{A}\) and \(\bm{B}\) are similar except for the circular shift of their columns, then \(\bm{A} = \bm{B} \, \bm{P}^{v}\) and:
\begin{equation}
\label{eqn_tdl_mat_y_isi_sep}
\bm{y_u} = \bm{H} \bm{x_u} + \bm{B} (\bm{x_{u-1}} - \bm{P}^{v} \bm{x_{u}}) + \bm{w_u}
\end{equation}
where \(\bm{P}^{v}\) denotes matrix power operation. Therefore, multiplication \(\bm{P}^{v} \bm{x_{u}}\) in (\ref{eqn_tdl_mat_y_isi_sep}) yields a circular-shift of elements in vector \(\bm{x_{u}}\) by \(v\) samples, i.e. the length of cyclic prefix.

Additionally, CIR matrix \(\bm{H}\) may be further decomposed into a sum of a circulant matrix \(\bar{\bm{H}}\) representing a time-invariant (time-averaged) part and an ICI matrix \(\bm{\Delta H}\) representing time-variability of the channel during single OFDM frame. Applying analogous operations to matrix \(\bm{B}\) yields:
\begin{equation}
\label{eqn_H_decom}
\bm{H} = \bar{\bm{H}} + \bm{\Delta H}
\end{equation}
\begin{equation}
\label{eqn_B_decom}
\bm{B} = \bar{\bm{B}} + \bm{\Delta B}
\end{equation}
where matrix \(\bar{\bm{B}}\) has zero-padded upper-triangular Toeplitz structure.

\section{Analysis of ISI Distortion in Frequency Domain}

Let \(\bm{F}\) be a normalized \(N \times N\) Fourier matrix with elements given by \([\bm{F}]_{n,m} = \frac{1}{\sqrt{N}} e^{-j 2 \pi n m / N}\), \(\bm{s_u} = \bm{F} \bm{x_u}\) a subcarrier mapped transmit (sent) data vector, \(\bm{r_u} = \bm{F} \bm{y_u}\) a received data vector in frequency domain and \(\bm{\eta_u} = \bm{F} \bm{w_u}\) a frequency domain noise. Moreover, denote a Channel Frequency Response (CFR) matrix as \(\bm{G} = \bm{F} \bm{H} \bm{F}^\dagger\) and an interference CFR matrix as \(\bm{\Phi} = \bm{F} \bm{B} \bm{F}^\dagger\), where superscript \(\dagger\) denotes Hermitian transpose. Using the time shift property of Fourier Transform, the DFT of circular-shift permutation matrix \(\bm{P}^v\) is given by:
\begin{equation}
\label{eqn_W}
\bm{W} = \bm{F} \bm{P}^{v} \bm{F} ^ \dagger = \text{diag} \Big( \Big[ 1, e^{-j 2 \pi v \frac{1}{N}}, \cdots , e^{-j 2 \pi v \frac{N-1}{N}} \Big] \Big)
\end{equation}

Applying Fourier Transform to both sides of (\ref{eqn_tdl_mat_y_isi_sep}) yields:
\begin{equation}
\label{eqn_freq_isi_sep}
\bm{r_u} = \bm{G} \bm{s_u} + \bm{\Phi} (\bm{s_{u-1}} - \bm{W} \bm{s_{u}}) + \bm{\eta_u}
\end{equation}

For the sake of the presented research, we assume that channel conditions change in time in a block fashion the way that CIR is constant during a given OFDM symbol, but it differs between consecutive symbols. It is a commonly accepted assumption in most practical scenarios \cite{Zhao08}. This implies that ICI parts in (\ref{eqn_H_decom}) and (\ref{eqn_B_decom}) are equal zeros, i.e. \(\bm{\Delta H} = \bm{0}\) and \(\bm{\Delta B} = \bm{0}\). Therefore, matrix \(\bm{H} = \bar{\bm{H}}\) is circulant, so \(\bm{G} = \bar{\bm{G}}\) is diagonal and contains eigenvalues of \(\bm{H}\). Matrix \(\bm{\Phi}\) is a DFT of upper-triangular Toeplitz matrix.

In \cite{Huckle93}, Huckle has derived a formula to compute DFT of Hermitian Toeplitz matrices. Using analogous reasoning, we may reformulate the equations for the upper-triangular Toeplitz matrix \(\bm{\Phi}\). Denote vector \(\bm{\rho} = [0, \hdots, 0, h(0,\tau), h(0,\tau-1), \hdots, h(0,v+1)]^T\) as the first row of matrix \(\bm{B}\) and \(\bm{\xi}\) as its DFT computed with reversed phase rotation, which is equivalent to IDFT operation:
\begin{equation}
\label{eqn_phi}
\bm{\xi} = \bm{F}^\dagger \bm{\rho}
\end{equation}
Then, the off-diagonal elements of matrix \(\bm{\Phi}\) can be calculated as:
\begin{equation}
\label{eqn_toep_dft}
[\bm{\Phi}]_{n,m} = \frac{1}{\sqrt{N}} \frac{1}{1 - w^{n-m}} ([\bm{\xi}]_m - [\bm{\xi}]_n)
\end{equation}
for \(n \neq m\), where \(w=e^{- j 2 \pi / N}\) is a \(N\)-th root of unity and \(n,m=0,\dots,N-1\) and \([\bm{\xi}]_n\) denotes the \(n\)-th element of vector \(\bm{\xi}\).

The diagonal part of \(\bm{\Phi}\) is given by eigenvalues of an optimal circulant preconditioner for matrix \(\bm{B}\) \cite{Chan88}. It may be calculated as an IDFT of Hadamard product, denoted by \(\circ\), between vector \(\bm{\rho}\) and a single period of reversed sawtooth function:
\begin{equation}
\label{eqn_toep_dft_diag}
\text{diag}(\bm{\Phi}) = \frac{1}{\sqrt{N}} \bm{F}^\dagger ([N,N-1,\dots,1]^T \circ \bm{\rho})
\end{equation}

Proof of (\ref{eqn_toep_dft}) and (\ref{eqn_toep_dft_diag}) can be found in the Appendix.

Matrix \(\bm{\Phi}\) without its diagonal, given by (\ref{eqn_toep_dft}), may be expressed as a Hadamard product between an envelope matrix \(\bm{E}\), dependent only on the DFT size \(N\) and invariant in time, and a matrix \(\bm{\Gamma}\) that depends on temporal channel conditions and a Power Delay Profile, which may differ between consecutive symbols. Therefore:
\begin{equation}
\label{eqn_toep_envelope}
[\bm{E}]_{n,m} = \begin{cases}
  \frac{1}{1 - w^{n-m}} & \quad \text{for } n \neq m \\
  0 & \quad \text{for } n = m
\end{cases}
\end{equation}
\begin{equation}
\label{eqn_toep_gamma}
[\bm{\Gamma}]_{n,m} =  [\bm{\xi}]_m - [\bm{\xi}]_n
\end{equation}
\begin{equation}
\label{eqn_hadamard}
\bm{\Phi} = \frac{1}{\sqrt{N}} \Big( \bm{E} \circ \bm{\Gamma} + \text{diag}\big(\bm{F}^\dagger ([N,N-1,\dots,1]^T \circ \bm{\rho})\big) \Big)
\end{equation} 

Note the following identity for any \(z \in \mathbb{R}\):
\begin{equation}
\label{eqn_tri_1over}
\frac{1}{1 - e ^ {j z}} = \frac{1}{2} \Big(1 + j \cot{\frac{z}{2}}\Big)
\end{equation}

Matrix \(\bm{E}\) is Hermitian-symmetric and \(\bm{\Gamma}\) is skew-symmetric, thus the energy of elements in both matrices and the result of their Hadamard product \(\bm{\Phi}\) is distributed symmetrically. We are interested in finding expected energies of elements in matrix \(\bm{\Phi}\) with respect to distance from its diagonal and its relation to undistorted subcarrier, namely, the relation between \(\mathbb{E}\{\|[\bm{G}]_{n,n} [\bm{s_u}]_n\|^2\}\) and \(\mathbb{E}\{\|[\bm{\Phi}]_{n,m} [\bm{s_u}]_m\|^2\}\).

Assume that transmit modulation data has uniform power \(\mathbb{E}\{\|[\bm{s_u}]_n\|^2\} = 1\). Then, the average power of undistorted subcarrier \(n\) is given by:
\begin{equation}
\label{eqn_Expec_H}
\begin{split}
\mathbb{E}\{\|[\bm{G}]_{n,n} [\bm{s_u}]_n\|^2\} & = \mathbb{E}\{\|[\bm{G}]_{n,n}\|^2\} \\ 
& = \frac{1}{N} \displaystyle\sum_{m=0}^{N-1} \big( \|h(n,m)\| \big) ^2
\end{split}
\end{equation}

Similarly, we derive a formula for the maximum power of ISI distortion generated by the \(m\)-th and affecting the \(n\)-th subcarrier, either from previous or current OFDM symbol. From (\ref{eqn_toep_envelope}), \([\bm{E}]_{n,m}\) is constant for given \(n\) and \(m\). Then, according to (\ref{eqn_hadamard}), for \(n \neq m\) we got:
\begin{equation}
\label{eqn_Expec_Phi}
\begin{split}
\mathbb{E}\{\|[\bm{\Phi}]_{n,m} [\bm{s_u}]_m\|^2\} & = \mathbb{E}\{\|[\bm{\Phi}]_{n,m}\|^2\} \\
& = \frac{1}{N} \|[\bm{E}]_{n,m}\|^2 \, \mathbb{E}\{\|[\bm{\Gamma}]_{n,m}\|^2\}
\end{split}
\end{equation}

In order to evaluate (\ref{eqn_Expec_Phi}), we expand \(\bm{\xi}\) in (\ref{eqn_toep_gamma}) to be expressed by elements of vector \(\bm{\rho}\):
\begin{equation}
\label{eqn_Delta_dft}
\begin{split}
[\bm{\Gamma}]_{n,m} & = \displaystyle\sum_{k=0}^{N-1} [\bm{\rho}]_k e ^{j 2 \pi m k / N} - \displaystyle\sum_{k=0}^{N-1} [\bm{\rho}]_k e ^{j 2 \pi n k / N} \\
& = \displaystyle\sum_{k=0}^{N-1} [\bm{\rho}]_k e^{j 2 \pi n k / N} \Big( 1 - e^{j 2 \pi (m-n) k / N} \Big)
\end{split}
\end{equation}

Moreover, the following trigonometric identity holds:
\begin{equation}
\label{eqn_tri_Gamma}
\big\|1 - e^{j z}\big\| = \big| 2 \sin{ \frac{z}{2} } \big|
\end{equation}

Using (\ref{eqn_Delta_dft}) and (\ref{eqn_tri_Gamma}), power of elements in matrix \(\bm{\Gamma}\) may be derived:
\begin{equation}
\label{eqn_Expec_Delta}
\begin{split}
\mathbb{E}\{\|[\bm{\Gamma}]_{n,m}\|^2\} = \frac{4}{N} \displaystyle\sum_{k=0}^{N-1} \Big( \|[\bm{\rho}]_k\| \cdot \Big| \sin{\frac{\pi (m-n) k} {N}} \Big|\Big) ^2
\end{split}
\end{equation}

Finally, incorporating (\ref{eqn_tri_1over}) and (\ref{eqn_Expec_Delta}) into (\ref{eqn_Expec_Phi}) yields:
\begin{equation}
\label{eqn_Expec_Phi_ext}
\begin{split}
&\mathbb{E}\{\|[\bm{\Phi}]_{n,m}\|^2\} = \\
&= \frac{\big(\cot\frac{\pi (m-n)}{N}\big)^2 + 1}{N} \displaystyle\sum_{k=0}^{N-1} \Big( \|[\bm{\rho}]_k\| \cdot \Big| \sin{\frac{\pi (m-n) k} {N}} \Big| \Big) ^2
\end{split}
\end{equation}

\begin{figure}[!t]
  \centering
  \includegraphics[width=2.8in]{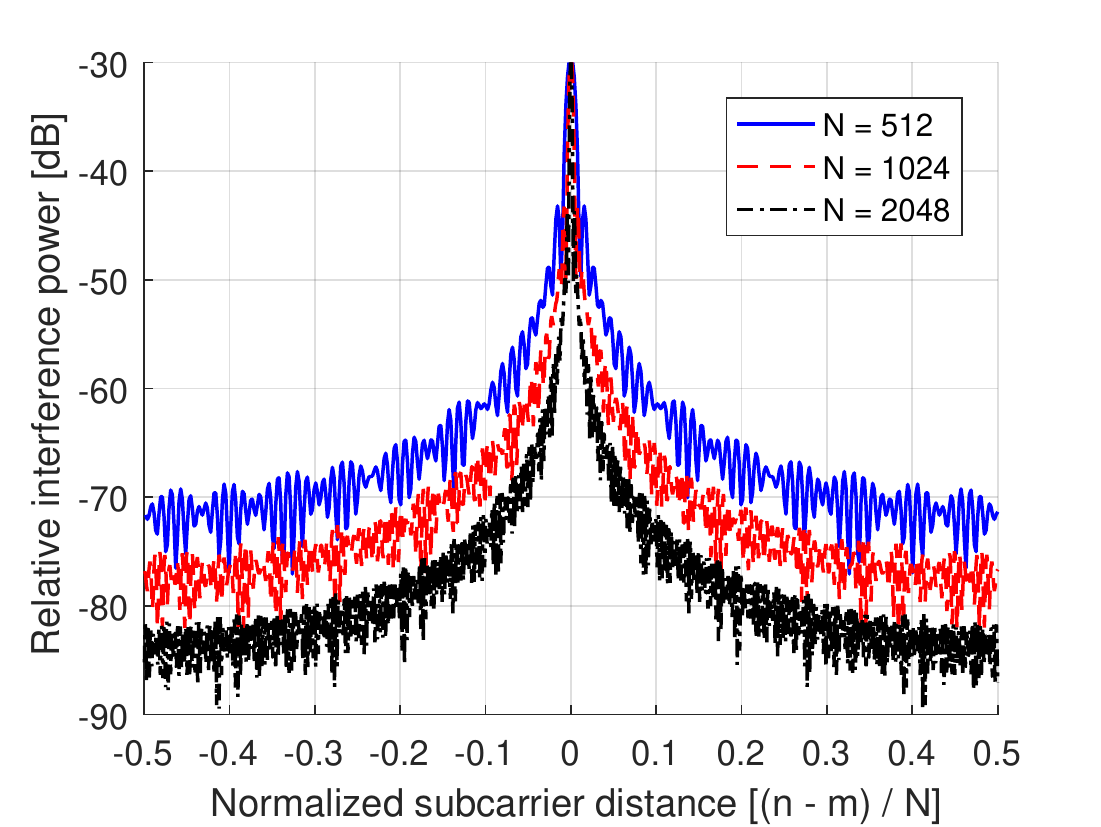}
  \caption{Power of interference matrix \(\bm{\Phi}\) relative to power of non-distorted subcarrier for various FFT sizes \(N\) and COST259 HT profile.}
  \label{fig1_rel_pwr}
\end{figure}

Equation (\ref{eqn_Expec_Phi_ext}) determines how significant is contribution of neighboring subcarriers to interference caused by insufficient cyclic prefix. The power of interference depends on FFT size, cyclic prefix, distance between subcarriers and a Power Delay Profile of the channel. When a duration of the longest multipath tap spreaded beyond the cyclic prefix is much smaller the the FFT size, which is always true in practical OFDM realizations, the average distortion power tends to decrease as the distance between subcarriers grows. As illustrated in Fig. \ref{fig1_rel_pwr}, the power of ISI is non-negligible only for the closest neighboring subcarriers, similarly as it was proven in case of Intercarrier Interference (ICI) inducted by high mobility \cite{Kim06}. 

The results consider COST259 Hilly Terrain (HT) multipath profile, which is a commonly accepted model to represent environments with very high delay spreads \cite{TR25.943}. 

\section{Numerical Experiments}

The major consequence of (\ref{eqn_Expec_Phi_ext}) is possibility to estimate or model the significant part of ISI in OFDM considering only the subcarriers concentrated in the neighborhood of the target subcarrier, analogously as in case of ICI \cite{Kim06}. 

We define a transformation of interference CFR matrix \(\bm{\Phi}\) to \(\bm{\tilde{\Phi}}\) that ignores the subcarriers with negligible energy outside the diagonal band \(b\), which results in a sparse structure:
\begin{equation}
\label{eqn_band_reduction}
[\bm{\tilde{\Phi}}]_{n,m} = \begin{cases}
  [\bm{\Phi}]_{n,m} &  |n - m| \le b \text{ or } |n - m| \ge N - b + 1 \\
  0 & \text{otherwise}
\end{cases}
\end{equation}

\subsection{Accuracy of ISI modeling with reduced CFR matrix}

\begin{figure}[!t]
  \centering
  \includegraphics[width=2.9in]{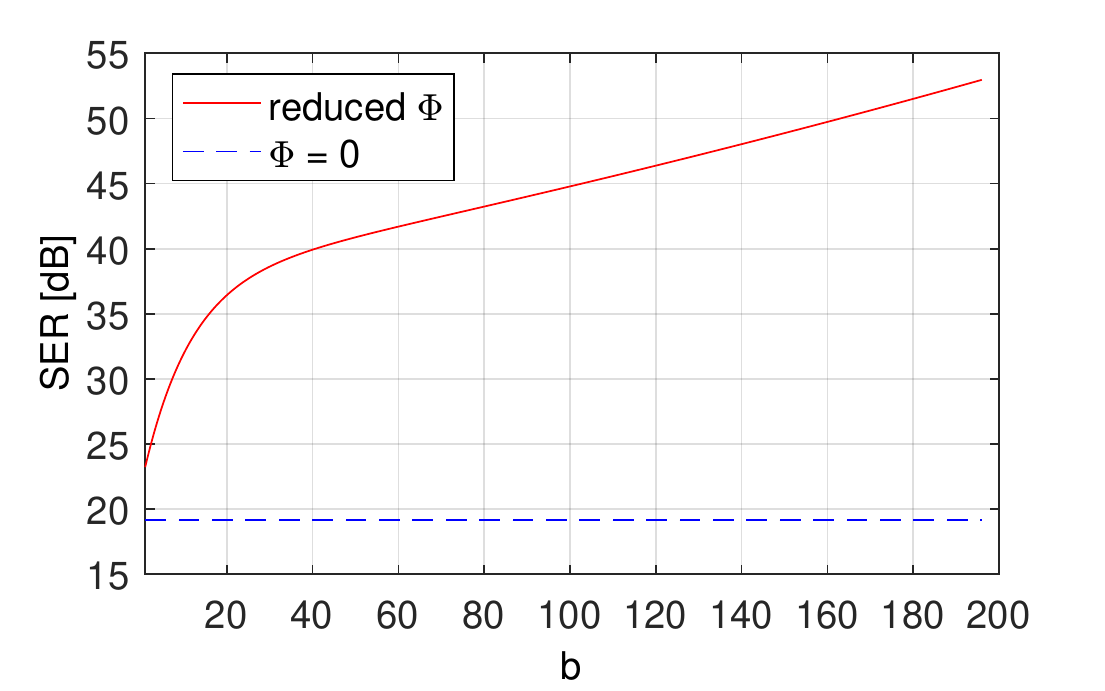}
  \caption{Accuracy of ISI modeling with reduced interference CFR matrix \(\bm{\tilde{\Phi}}\) relative to non-reduced approach. The results are averaged over \(10^4\) symbols of 5 MHz LTE (\(N=512\)) and COST259 HT profile.}
  \label{fig2_accuracy}
\end{figure}

For the purpose of simulation, an OFDM transmitter based on the 3GPP Long Term Evolution (LTE) standard with normal cyclic prefix and a full bandwidth coverage is considered \cite{TS36.104}. The randomly generated stream of 16-QAM IQ symbols is OFDM modulated and passed in parallel through multipath fading channel models with both reduced \(\bm{\tilde{\Phi}}\) and complete \(\bm{\Phi}\) CFR, according to (\ref{eqn_band_reduction}). Then, the resulting data is OFDM demodulated and compared in term of Signal-to-Error Ratio (SER) defined accordingly:
\begin{equation}
\label{eqn_sim_snr}
SER = 20 \cdot \log_{10}{\frac{\|\bm{\tilde{r}}\|} {\|\bm{\tilde{r}} - \bm{r}\|}}
\end{equation}
where \(\bm{\tilde{r}}\) and \(\bm{r}\) are the responses of the channel with reduced and complete CFR, respectively. The simulation exploits COST259 HT multipath profile with 5 Hz maximum Doppler frequency. The results shown in Fig. \ref{fig2_accuracy} indicate that using even small values of \(b\) is sufficient to increase accuracy above 12 dB comparing to naive block-fading model with \(\bm{\Phi} = \bm{0}\).

\subsection{Normality test of residual ISI}

Reduction of CFR interference matrix according to (\ref{eqn_band_reduction}) results in residual pseudo-noise signal produced by the elements of CFR matrix \(\bm{\Phi} - \bm{\tilde{\Phi}}\). A Monte-Carlo simulation was accomplished to measure a goodness of fit to normal distribution of the residual signal using Jarque-Bera (JB) test statistics \cite{Jarque87}. The test sample is calculated as \(\sum_{k=0}^{N-1} [\bm{\Phi} - \bm{\tilde{\Phi}}]_{m,k} [\bm{s}]_k\), where the elements of vector \(\bm{s}\) are randomly chosen from 16-QAM alphabet for each sample. Calculated \(p\)-values are averaged over multiple realizations of a channel with HT multipath profile. The results are shown in Table \ref{table_normality}. Values below 0.05 reject the null-hypothesis that test sample may come from normal distribution, whereas higher values indicate a better match. The results draw a conclusion that distribution of the residual signal is closest to Gaussian for \(b\) values between \(N / 16\) and \(N / 8\). The interval is a trade off between the Central Limit Theorem and a presence of terms with dominant energy.

\setlength\tabcolsep{7pt}
\begin{table}[!t]
\renewcommand{\arraystretch}{1.2}
\caption{Obtained \(p\)-values from JB normality test statistics. Values computed with sample size of \(10^5\) and \(10^3\) channel realizations.}
\label{table_normality} 
\centering
\begin{tabular}{|c|c|c|c|c|}
\hline \backslashbox{\(b\)}{\(N\)} 
           &      128 &      256 &      512 &     1024 \\ 
\hline   0 & 0.000313 & 0.000105 & 0.053446 & 0.030402 \\ 
\hline   4 & 0.106961 & 0.049981 & 0.208508 & 0.093885 \\ 
\hline   8 & 0.321353 & 0.103199 & 0.258352 & 0.140079 \\ 
\hline  16 & 0.319279 & 0.376248 & 0.357483 & 0.319884 \\ 
\hline  32 & 0.107915 & 0.534479 & 0.402031 & 0.402015 \\ 
\hline  64 & 0.000000 & 0.515057 & 0.571853 & 0.487761 \\ 
\hline 128 & -        & 0.000000 & 0.427937 & 0.378077 \\ 
\hline 256 & -        & -        & 0.000000 & 0.316982 \\ 
\hline
\end{tabular}
\end{table}

\subsection{Application to channel modeling - Complexity}

\begin{figure}[!t]
  \centering
  \includegraphics[width=2.9in]{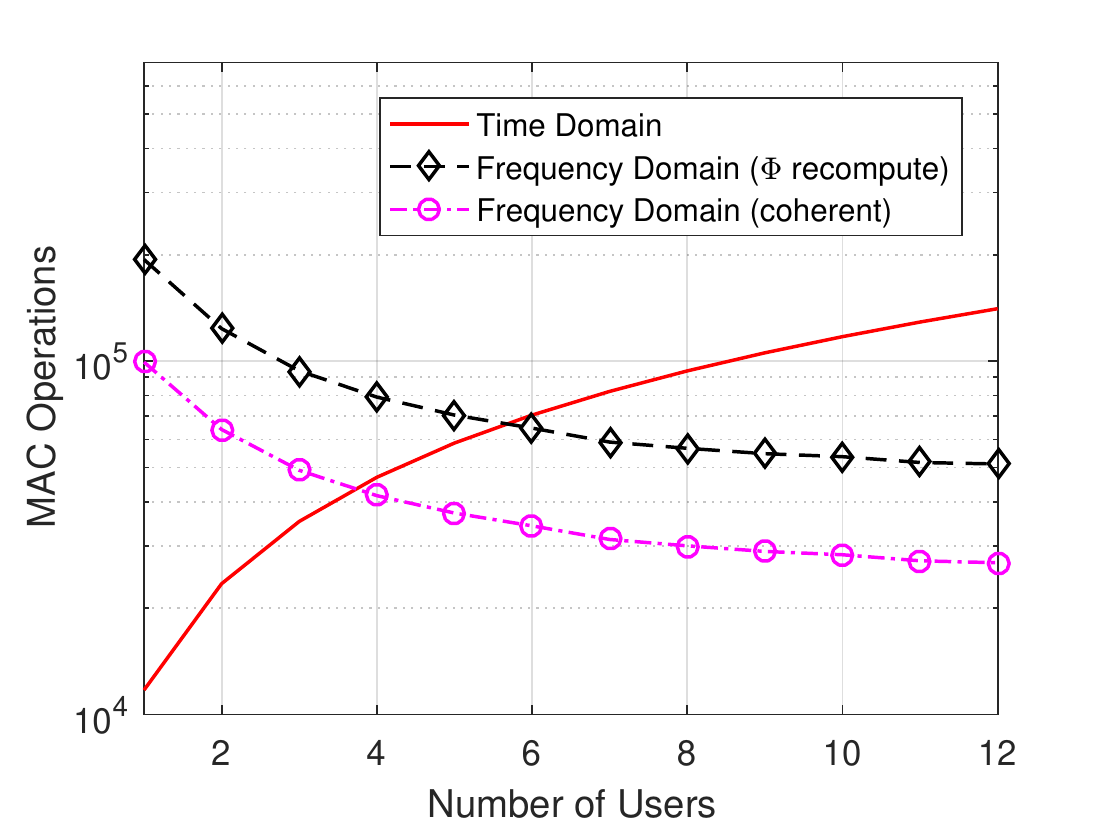}
  \caption{Complexity of frequency domain ISI simulation comparing to time domain Tapped Delay Line approach expressed in number of MAC operations. Generated for 5 MHz LTE (\(N=512\)), \(b = 16\) and COST259 HT profile.}
  \label{fig3_complexity}
\end{figure}

One application of the reported research is an area of multi-user frequency domain channel simulation. To evaluate the performance, we will use the same methodology for complexity metrics as used in \cite{Cisek18}. Two cases are compared against the time domain TDL model. The first one assumes that the coherence time of the channel is much longer than the OFDM symbol duration and the same CFR matrix \(\bm{\tilde{\Phi}}\) is used across many symbols. The second one assumes that \(\bm{\tilde{\Phi}}\) must be recalculated for every symbol. Numerical results were obtained by counting required complex Multiply and Accumulate (MAC) operations. Fig. \ref{fig3_complexity} shows that for small values of \(b\), frequency domain modeling gains advantage when the bandwidth is shared between multiple users.

\section{Conclusion}

The new framework presented in this paper describes the relation between OFDM subcarriers in frequency domain under ISI distortion. It was shown that the strongest ISI contribution to a given subcarrier comes from the neighboring subcarriers and it decays as the distance between subcarriers grows, similarly as in case of ICI. Due to this fact, frequency-domain receiver or transmit precoder may focus on compensation of the banded diagonal part of the CFR ISI matrix \(\bm{\Phi}\) to provide significant ISI distortion reduction. Numerical experiment has shown that the residual ISI for a given subcarrier, resulting from matrix \(\bm{\Phi}\) reduction to a banded one, has a distribution close to the normal for moderate-low band sizes. As a consequence, the residual ISI may be treated as additional Gaussian noise, correlated between neighboring subcarriers. 

A possible framework application to frequency domain wideband channel simulation was also investigated. Simulation time using the model presented in the paper is shorter than the time domain TDL model when a number of simulated users is high and reduction of CFR matrix is used. This fact justifies utilization of the model in the area of multi-user channel emulation.

\appendix[Proof of Equations (\ref{eqn_toep_dft}) and (\ref{eqn_toep_dft_diag})]

The purpose of this appendix is to derive a general formula for fast DFT of upper-triangular Toeplitz matrix \(\bm{B}\). However, a reader should keep in mind that the specific application presented in this paper considers a sparse zero-padded upper-triangular Toeplitz structure. 

Start from a standard DFT matrix equation:
\begin{equation}
\label{eqn_appen_phi}
\bm{\Phi} = \bm{F} \bm{B} \bm{F} ^ \dagger
\end{equation}

Multiplying the matrices on the left-hand side yields:
\begin{equation}
\label{eqn_appen_FB}
[\bm{F B}]_{n,m} = \frac{1}{\sqrt{N}} \displaystyle\sum_{k=0}^{m} [\bm{\rho}]_k \, w^{(m-k) n}
\end{equation}
where \(w=e^{- j 2 \pi / N}\) and vector \(\bm{\rho}\) is the first row of Toeplitz matrix \(\bm{B}\). Then, the elements of matrix \(\bm{\Phi}\) are given by:
\begin{equation}
\label{eqn_appen_phi_elem}
[\bm{\Phi}]_{n,m} = \frac{1}{N} \displaystyle\sum_{l=0}^{N-1} \Big( \displaystyle\sum_{k=0}^{m} [\bm{\rho}]_k \, w^{(m-k) n} \Big) w^{-ml}
\end{equation}

By unwrapping the sums in (\ref{eqn_appen_phi_elem}) and grouping with respect to elements of \(\bm{\rho}\) we got:
\begin{equation}
\label{eqn_appen_phi_elem2}
\begin{split}
[\bm{\Phi}]_{n,m} & = \frac{1}{N} \displaystyle \sum_{k=0}^{N-1} [\bm{\rho}]_k \Big( \displaystyle \sum_{l=k}^{N-1} w^{(l-k)n - lm} \Big) \\
& = \frac{1}{N} \displaystyle \sum_{k=0}^{N-1} [\bm{\rho}]_k w^{-km} \Big( \displaystyle \sum_{l=0}^{N-k-1} w^{l(n - m)} \Big)
\end{split}
\end{equation}

Consider (\ref{eqn_appen_phi_elem2}) in two cases:

\paragraph{\(n = m\)}

The exponents in the inside sum are equal 0 for all \(k\), thus:
\begin{equation}
\label{eqn_appen_phi_neqm}
[\bm{\Phi}]_{n,n} = \frac{1}{N} \displaystyle \sum_{k=0}^{N-1} (N - k) \, [\bm{\rho}]_k w^{-kn}
\end{equation} 
which is equivalent to (\ref{eqn_toep_dft_diag}).

\paragraph{\(n \neq m\)}

Expand the inside sum using the formula for the finite sum of geometric series:
\begin{equation}
\label{eqn_appen_phi_nneqm}
\begin{split}
[\bm{\Phi}]_{n,m} & =\frac{1}{N}  \displaystyle \sum_{k=0}^{N-1} [\bm{\rho}]_k w^{-km} \frac{1 - w ^ {(n-m)(M-k)}}{1 - w^{n-m}} \\
& = \frac{1}{N} \frac{1}{1-w^{n-m}} \Big( \displaystyle \sum_{k=0}^{N-1} [\bm{\rho}]_k w^{-km} - \displaystyle \sum_{k=0}^{N-1} [\bm{\rho}]_k w^{-kn} \Big) \\
& = \frac{1}{\sqrt{N}} \frac{1}{1-w^{n-m}} ([\bm{\xi}]_m - [\bm{\xi}]_n)
\end{split}
\end{equation}
what demonstrates (\ref{eqn_toep_dft}). \(\blacksquare\)

\small
\section*{Acknowledgment}

G. Cisek was supported in this work by AGH University of Science and Technology Dean’s Grant while T. Zieliński was financed by AGH University of Science and Technology contract no 11.11.230.018.

\end{document}